\documentstyle[prl,aps,preprint]{revtex}
\draft
\begin{document}
\title{Coulomb drag in double quantum wells with a perpendicular magnetic
field}
\author{M.W. Wu, H.L. Cui, and N.J.M. Horing}
\address{Department of Physics and Engineering Physics, Stevens Institute
of Technology, Hoboken, NJ 07030}
\maketitle

\begin{abstract}
Momentum transfer due to electron-electron interaction (Coulomb drag)
between two quantum wells, separated by a distance $d$, in
the presence of a perpendicular magnetic field, is studied at low
temperatures. We find besides the well known Shubnikov-de Haas oscillations,
which also appear in the drag effect, the momentum transfer is
markedly enhanced by the magnetic field.
\end{abstract}


Coulomb drag effect of double quantum wells has recently attracted much
experimental\cite{sl,gr1,si,gr2} and
theoretical\cite{la,bo,so,ro,to,ma,du,j,z,cu,shi,fle,sw,f,fl} attention.
This is because, not withstanding
the fact that the effects of electron-electron collisions only have indirect
consequences for transport properties of single isolated quantum wells as
they conserve total momentum and cannot transfer or relax it,
the drag effect of double quantum wells is unique in that it provides an
opportunity to directly measure electron-electron interaction through a
transport measurement where momentum is transferred from one well to the
other. Consider two quantum wells containing electrons so close to each other
that the electrons in the two respective wells experience the Coulomb forces
originating from the other well and yet far enough away from each other that
direct charge transfer between the two wells is not possible. Thus, if a
current $J$ is driven through one of the wells, then an induced current is
dragged in the other well. Alternatively, if no current is allowed to flow in
the other well, an electric field $E$ is induced. The so called
transresistivity $\rho$, which describes the momentum transfer between the
two wells, is hereupon defined as
\begin{equation}
\rho=\frac{E}{J}\;.
\end{equation}
To our knowledge, almost all the works on this subject are focused on
screening effects associated with plasmon modes
(see, {\em eg.}\cite{sw,f,fl}), or on deviations of the
transresistivity from $T^2$ law at low temperature (from zero to several
Kelvin degrees) (see, {\em eg.}\cite{gr2,to,fle}). Magnetic field effects
on Coulomb drag in planar quantum wells have, to a large extent\cite{tso},
escaped attention. It is known that in the classical limit, the magnetic
field does not affect the Coulomb drag effect. Nevertheless, for higher
magnetic fields,  when Landau quantization becomes important, the situation
is quite different. Since the mechanism for Coulomb drag is
carrier-carrier scattering, the transresistivity
is proportional to the square of the effective interaction
between the two wells. The available phase space for electron-electron
scattering is modified drastically by the magnetic field due to the
formation of Landau levels, consequently affecting Coulomb drag.

The purpose of this letter is to elucidate
the significance of the magnetic field, which is applied perpendicular to
the quantum wells, in regard to the Coulomb drag effect in double quantum
wells at low temperature. We also include the role of dynamic screening
of the Coulomb interactions between two quantum wells. As may be expected,
we find the well known Shubnikov-de Haas oscillations, which feature
prominently in the drag effect when temperature is sufficiently low.
Our analysis further shows that the Coulomb drag
effect is remarkably enhanced by high magnetic field at low temperature.

The transresistivity, which can be calculated from either the Boltzmann
equation\cite{gr2,j}, the momentum balance equation method\cite{to,cu}, or,
very recently, the  Kubo linear-response formula\cite{f,fl}, is given by
\begin{equation}
\label{rho}
\rho=\frac{1}{4\pi^2n_1n_2e^2}\int_0^\infty dq q^3\int_{-\infty}^\infty
\frac{d\omega}{T}\left|\frac{v(q)e^{-bq}}{\varepsilon(q,\omega)}\right|^2
\left[-n^\prime
\left(\frac{\omega}{T}\right)\right]\Pi_2^{(1)}(q,\omega)
\Pi_2^{(2)}(q,\omega)\;.
\end{equation}
In this equation, $n_1$ ($n_2$) stands for the sheet density of first
(second) quantum well. $n(x)=[\exp(x)-1]^{-1}$ is the Bose distribution
function and $n^\prime(x)=\frac{d}{dx}n(x)$. $v(q)=2\pi e^2/\kappa q$ is the
bare 2D Coulomb interaction, with $\kappa$ being the background dielectric
constant. The dynamic screening of the Coulomb interaction, in random
phase approximation, reads
\begin{equation}
\label{epsilon}
\varepsilon(q,\omega)=[1-v(q)\Pi^{(1)}(q,\omega)]
[1-v(q)\Pi^{(2)}(q,\omega)]
-v(q)^2e^{-2bq}\Pi^{(1)}(q,\omega)\Pi^{(2)}(q,\omega)\;.
\end{equation}
Here $\Pi^{(j)}(q,\omega)$ is the electron density-density correlation
function of $j$th quantum well, with $\Pi^{(j)}_1(q,\omega)$ ($\Pi^{(j)}_2
(q,\omega)$) denoting the real (imaginary) part of it. In the presence of
magnetic field, it is given by\cite{ting}
\begin{equation}
\Pi^{(j)}(q,\omega)=\frac{1}{\pi\alpha^2}\sum_{nn^\prime}C_{nn^\prime}(x)
\Pi^{(j)}(n,n^\prime,\omega)\;,
\end{equation}
where
\begin{equation}
C_{nn^\prime}(x)=\frac{m_2!}{m_1!}x^{m_1-m_2}e^{-x}[\mbox{L}_{m_2}^{m_1
-m_2}(x)]^2
\end{equation}
with $m_1=\max (n,n^\prime)$ and $m_2=\min (n,n^\prime)$. $\mbox{L}_m
^{m^\prime}(x)$ is the associated Laguerre polynomial. $\alpha$
is the radius of the ground cyclotron orbit, given by
$\alpha=(eB)^{-1/2}$ with $B$ denoting the magnetic field.
$x=\alpha^2q^2/2$. The quantity $\Pi^{(j)}(n,n^\prime,\omega)$ can be
expressed as
\begin{eqnarray}
\mbox{Re}\Pi^{(j)}(n,n^\prime,\omega)&=&-\frac{1}{\pi}\int_{-\infty}^{\infty}
dz[f_j(z)-f_j(z+\omega)]\mbox{Re}G^{(j)}_n(z+\omega)
\mbox{Im}G^{(j)}_{n^\prime}(z)\;,\\
\mbox{Im}\Pi^{(j)}(n,n^\prime,\omega)&=&-\frac{1}{\pi}\int_{-\infty}^{\infty}
dz[f_j(z)-f_j(z+\omega)]\mbox{Im}G^{(j)}_n(z+\omega)
\mbox{Im}G^{(j)}_{n^\prime}(z)\;.
\end{eqnarray}
In these equations, $f_j(z)$ denotes the Fermi distribution function of $j$th
well, given by $\{\exp[(z-\mu_j)/T]+1\}^{-1}$ with $\mu_j$ denoting
chemical potential, which is determined, for total electron sheet density
$n_j$, by the relation
\begin{equation}
\label{n}
\frac{1}{\pi^2\alpha^2}\sum_{n=0}^{\infty}\int d\omega f_j(\omega)\mbox{Im}
G^{(j)}_n(\omega)=n_j\;.
\end{equation}
$G^{(j)}_n(z)$, the Green function for the $n$th Landau level of
$j$th well, can be expressed,
in the self-consistent Born approximation\cite{ando,ando1}, as
\begin{equation}
G^{(j)}_n(\omega)=\frac{2}{\Gamma_j^2}[\omega-\varepsilon_n-\sqrt{(\omega-
\varepsilon_n)^2-\Gamma_j^2}]\;,
\end{equation}
with $\Gamma_j^2=2\omega_c/\pi\tau^{(j)}_0$ and $\tau^{(j)}_0$ is
the electron transport
lifetime in the absence of a magnetic field at zero temperature, related to
the mobility $\mu^{(j)}_0$ under the same conditions by $\tau^{(j)}_0
=m\mu^{(j)}_0/e$. $\varepsilon_n=(n+1/2)\omega_c$ is the Landau level
energy with $\omega_c=eB/m$.

Based on Eqs.\ (\ref{rho}) and (\ref{n}) we can calculate the
transresistivity, as a function of magnetic field at low temperatures, to
examine the influence of Landau quantization of magnetic field on the
Coulomb drag effect. The magnetic field considered in our calculation is
taken large enough so that
the Landau level energy $\hbar\omega_c$ is larger or at least comparable to
$k_BT$, and therefore quantum effects are important. Parameters
pertaining to a GaAs-AlGaAs-GaAs double quantum well structure are used,
with electron effective mass $m=0.07m_e$ ($m_e$ is the free electron mass),
barrier thickness $b=200$\AA ~and the background dielectric constant
$\kappa=12.9$. We suppose that both quantum wells share the same sheet
density $n_1=n_2=10^{15}$\ m$^{-2}$, and the mobility at zero temperature is
$\mu_0^{(1)}=\mu_0^{(2)}\equiv\mu_0=25$\ m/Vs.

We first discuss the collective modes of the coupled electron gas. They are
given by the zeros of the real part of the dielectric function
Eq.\ (\ref{epsilon}), Re$[\varepsilon(q,\omega(q))]=0$ at finite temperature,
with the imaginary part describing the damping. As in the nonmagnetic
field case\cite{das}, there are two collective modes,
as shown in Fig.\ 1 in which we plot the integrand of Eq.\ (\ref{rho})
versus $\omega/T$ for different values of $q$ normalized by $\alpha$,
when $T=3$\ K (Fig.\ 1(a)) and 10\ K (Fig.\ 1(b)). The magnetic field in
these figures is 1\ T.  From the figure we can see that as $q$ increases, the
two modes both increase in frequency $\omega(q)$ and become closer.
On the other hand, as $q$ goes to zero, one mode $\omega(q)$ also tends to
zero, which is the so called acoustic mode, and the other is in the nature
of an optical mode with $\omega(q)\not=0$. We point out here that this
feature of the optical mode in the presence of the magnetic field is
quite different from the nonmagnetic case, as in that circumstance, the
optical mode also tends to zero as $q\to 0$, with its small $q$ limiting
behavior at zero temperature being $q^{1/2}$\cite{fl,das}. This is because,
in the absence of magnetic field, the collective excitation reflects the
nature of 2D electron gases, whereas in the presence of the field,
this 2D freedom is further restricted by Landau quantization.

In Fig.\ 2 we plot the numerically calculated transresistivity $\rho$, which
is normalized by $\rho_0$\cite{comment} as a function of magnetic field $B$
at different temperatures $T=3$, 4, 6.5, 10, and 15\ K. We include ten Landau
levels in our computation.  In order to exhibit the screening effect, we plot
the transresistivity both with and without screening in Fig.\ 2(a) and (b)
respectively. From both of these figures we can see that the magnetic field
has a strong effect on the transresistivity, especially at low temperatures.
We note, from the figures for $T$ below about 4\ K, the transresistivity
exhibits distinctive Shubnikov-de Haas oscillations as the magnetic field
increases from 0.5\ T to 2.5\ T  (with the filling factor $\nu$ decreasing
from 8.3 to 1.7). It is also clear from the figures
that the minima for the
curve of $T=3$\ K from the right to the left correspond to the filling
factors $\nu$=2, 4, 6, 8, respectively, which means the Fermi level
is between the first and the second, second and third, $\cdots$
{\em etc.}, Landau levels. Hence the
available phase space is greatly reduced, thus reducing the transresistivity.
As the magnetic field continues to increase, eventually all the electrons
are accommodated in the first Landau level, the transresistivity is markedly
enhanced, becoming an order of magnitude larger than that in the low field
case. However, when temperature becomes higher, the oscillations are washed
out due to thermal fluctuation and the magnetic enhancement of the drag
effect becomes moderate as $B$ increases.

We also note from the figures that screening is more effective for low
temperatures than for higher ones. This can be well illustrated
from the fact that in Fig.\ 2(a) one can see that when electron screening
effect is included, for any magnetic field, the drag effect always
increases as temperature rises. However in Fig.\ 2(b) one can see, especially
in the high magnetic field regime ({\em eg.} $B>2.5$\ T), that the lower the
temperature, the higher the transresistivity. This implies that screening
reduces the drag effect at low temperature faster than at high
temperature. This is readily understood since the degree of
degeneracy is higher at low temperatures, so that
the screening of electrons is more effective than at high temperature case.

In summary, we have studied the effects of a quantizing magnetic field
on the momentum transfer due to electron-electron Coulomb interaction
between two spatially separated quantum wells. We find that, at low
temperatures, the magnetic field not only induces Shubnikov-de Haas
oscillations, but it also markedly enhances  momentum transfer due to
electron-electron interactions.

\acknowledgements

One of the authors (MWW) would like to thank Mr. X.G. Feng, for providing
information about his pertinent experimental work in progress.
This research is
supported by U.S. Office Naval Research (Contract No. N66001-95-M-3472),
and the U.S. Army Research Office (Contract NO. DAAH04-94-G-0413).

\bigskip
\bigskip

\begin{center}
Fig.\ 1. Transresistivity (integrand of Eq.\ (\ref{rho})) as a function of
energy transfer $\omega$ scaled by temperature $T$, for various
wave vectors $q$ normalized by $\alpha$. (a): $T=$3\ K; (b): $T=10$\ K.
\end{center}

\bigskip
\bigskip

\begin{center}
Fig.\ 2. Transresistivity $\rho$ normalized by $\rho_0$ is plotted as a
function of magnetic field $B$ for various temperatures $T=3$, 4, 6.5,
10, and 15\ K. (a): with the interlayer screening effects by $\varepsilon(q,
\omega)$ in Eq.\ (\ref{epsilon}); (b): without screening effects {\em ie.}
$\varepsilon(q,\omega)=1$.
\end{center}

\begin{references}
\bibitem{comment} $\rho_0$ is defined from the overall factor of
Eq.\ (\ref{rho}): $e^2T^2/(\pi^6n_1^2\alpha^6\varepsilon_0^2
\kappa^2\Gamma^4)$ with $T=1$\ K and $B=1$\ T.
\end{references}
\end{document}